\documentclass[prl,twocolumn]{revtex4}

\usepackage{graphicx}

\def\r{\rangle}

\newcommand{\be}{\begin{eqnarray}}
\newcommand{\ee}{\end{eqnarray}}

\begin{document}

\title{Towards quantum superpositions of a mirror}

\author{William Marshall$^{1,2}$, Christoph Simon$^1$, Roger Penrose$^{3,4}$ and Dik Bouwmeester$^{1,2}$}
\address {$^1$Department of Physics, University of Oxford, Oxford OX1 3PU, United Kingdom\\ $^2$
Department of Physics, University of California, Santa Barbara, CA
93106\\
$^3$ Center for Gravitational Physics and Geometry, The Pennsylvania
State University, University
Park, PA 16802\\
$^4$ Department of Mathematics, University of Oxford, Oxford OX1
3LB, United Kingdom}

\date{\today}

\begin{abstract}
We propose a scheme for creating quantum superposition states
involving of order $10^{14}$ atoms via the interaction of a single
photon with a tiny mirror. This mirror, mounted on a high-quality
mechanical oscillator, is part of a high-finesse optical cavity
which forms one arm of a Michelson interferometer. By observing
the interference of the photon only, one can study the creation
and decoherence of superpositions involving the mirror. All
experimental requirements appear to be within reach of current
technology.
\end{abstract}

\maketitle

\bigskip

In $1935$ Schr\"{o}dinger pointed out that according to quantum
mechanics even macroscopic systems can be in superposition states
\cite{schroedinger}. The quantum interference effects are expected
to be hard to detect due to environment induced decoherence
\cite{decoherence}. Nevertheless there have been proposals on how
to create and observe macroscopic superpositions in various
systems \cite{bec,bose,armour}, and experiments demonstrating
superposition states of superconducting devices \cite{squidexp}
and fullerene molecules \cite{c60}. One long-term motivation for
this kind of experiment is the question of whether unconventional
decoherence processes such as gravitationally induced decoherence
or spontaneous wave-function collapse \cite{collapse,penrose}
occur.

Here we present a scheme that is close in spirit to
Schr\"{o}dinger's original discussion. A small quantum system (a
photon) is coupled to a large system (a mirror) in order to create
a macroscopic superposition. The basic principle of the experiment
as described in Ref. \cite{penrose} grew out of discussions in
1997 \cite{schmiedmayer}. It consists of a Michelson
interferometer in which one arm has a tiny moveable mirror. The
radiation pressure of a single photon displaces the tiny mirror.
The initial superposition of the photon being in either arm causes
the system to evolve into a superposition of states corresponding
to two distinct locations of the mirror. In the present proposal a
high-finesse cavity is used to enhance the interaction between the
photon and the mirror. The observed interference of the photon
allows one to study the creation of coherent superposition states
periodic with the motion of the mirror. We perform a detailed
analysis of the requirements for such an experiment.

\begin{figure}
\includegraphics[width=0.7 \columnwidth]{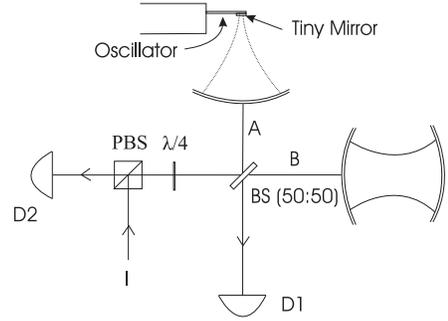}
\caption{The proposed setup: a Michelson interferometer for a
single photon, where in each arm there is a high-finesse cavity.
The cavity in arm A has a very small end mirror mounted on a
micro-mechanical oscillator. The single photon comes in through
I. If the photon is in arm A, the motion of the small mirror is
affected by its radiation pressure. The photon later leaks out of
either cavity and is detected at D1 or D2.} \label{setup}
\end{figure}

 The proposed setup is shown in Fig. 1. In the
cavity in arm A one of the mirrors is very small and attached to a
micromechanical oscillator, similar to the cantilevers in atomic
force microscopes. While the photon is in the cavity, it exerts
radiation pressure on the mirror. Under the conditions that the
oscillator period is much longer than the photon roundtrip time,
and the amplitude of the mirror's motion is small compared to the
cavity length, the system can be described by a Hamiltonian
\cite{law} \be H=\hbar \omega_c a^{\dagger} a + \hbar \omega_m
b^{\dagger} b - \hbar g a^{\dagger} a (b+b^{\dagger}), \ee where
$\omega_c$ is the frequency of the photon in the cavity for the
(empty) cavity length $L$, $a$ is the creation operator for the
photon, $\omega_m$ and $b$ are the frequency and phonon creation
operator for the center of mass motion of the mirror, and the
coupling constant is $g=\frac{\omega_c}{L}\sqrt{\frac{\hbar}{2M
\omega_m}}$, where $M$ is the mass of the mirror. To start with,
let us suppose that initially the photon is in a superposition of
being in either arm $A$ or $B$, and the mirror is in a coherent
state $|\beta \r = e^{-|\beta|^{2}/2} \sum \limits_{n=0}^{\infty}
\frac{\beta^{n}}{\sqrt{n!}}|n\r$, where $|n\r$ are the eigenstates
of the harmonic oscillator. Then the initial state is
$|\psi(0)\r=\frac{1}{\sqrt{2}}(|0\r_{A} |1\r_{B} + |1\r_{A}
|0\r_{B})|\beta\r$. After a time $t$ the state of the system will
be given by \cite{mancini} \be
|\psi(t)\r=\frac{1}{\sqrt{2}}e^{-i\omega_c t}(|0\r_A |1\r_B |\beta
e^{-i \omega_m t}\r + \nonumber\\ e^{i\kappa^2(\omega_m t-\sin
\omega_m t)}|1\r_A |0\r_B  |\beta e^{-i \omega_m t}+
\kappa(1-e^{-i \omega_m t})\r ), \label{state} \ee where
$\kappa=g/\omega_m$. In the second term on the r.h.s. the motion
of the mirror is altered by the radiation pressure of the photon
in cavity $A$. The parameter $\kappa$ quantifies the displacement
of the mirror in units of the size of the coherent state
wavepacket. In the presence of the photon the mirror oscillates
around a new equilibrium position determined by the driving force.

The maximum possible interference visibility for the photon is
given by twice the modulus of the off-diagonal element of the
photon's reduced density matrix. By tracing over the mirror one
finds from Eq. (\ref{state}) that the off-diagonal element has the
form \be
\frac{1}{2} e^{-\kappa^2(1-\cos \omega_m t)} e^{i\kappa^2(\omega_m
t-\sin \omega_m t)+i \kappa \mbox{\small Im}[\beta (1-e^{i\omega_m
t})]}
\label{coherence} \ee where Im denotes the imaginary part. The
first factor is the modulus, reaching a minimum after
half a period at $t=\pi/\omega_m$, when the mirror is at its
maximum displacement. The second factor gives the phase, which is
identical to that obtained classically due to the varying length
of the cavity.
For general $t$ the phase depends on
$\beta$, i.e. the initial conditions of the mirror. However, the
effect of $\beta$  averages out after every full
period.

In the absence of decoherence, after a full period, the system is
in the state $\frac{1}{\sqrt{2}}(|0\r_A |1\r_B + e^{i\kappa^2 2
\pi} |1\r_A |0\r_B) |\beta\r$, such that the mirror is again
disentangled from the photon. Full interference can be observed if
the photon is detected at this time. If the environment of the
mirror ``remembers'' that the mirror has moved, then, even after a
full period, the photon will still be entangled with the mirror's
environment, and thus the interference for the photon will be
reduced. Therefore the setup can be used to measure the
decoherence of the mirror.

In practice the mirror will be in a thermal state, which
can be written as a mixture of coherent states $|\beta\r$ with a
Gaussian probability distribution $(1/\pi
\bar{n})e^{-|\beta|^2/\bar{n}}$, where $\bar{n}=1/(e^{\hbar
\omega_m/kT}-1)$ is the mean
thermal number of excitations.
In order to determine the expected
interference visibility of the photon at a time $t$ for an initial
mirror state which is thermal, one has to average the
off-diagonal element Eq. (\ref{coherence}) over $\beta$ with this
distribution. The result is \be \frac{1}{2}
e^{-\kappa^2(2\bar{n}+1)(1-\cos \omega_m t)} e^{i\kappa^2(\omega_m
t-\sin \omega_m t)}. \label{thermal} \ee As a consequence of the
averaging of the $\beta$-dependent phase in Eq.
({\ref{coherence}}), the off-diagonal element now decays on a
timescale $1/(\kappa \omega_m \sqrt{\bar{n}})$ after $t=0$, i.e.
very fast for the experimentally relevant case of $\kappa \sim 1$
and large $\bar{n}$. However, remarkably, it still exhibits a
revival \cite{bose} at $t=2\pi/\omega_m$, when photon and mirror
become disentangled and the phase in Eq. (\ref{coherence}) is
independent of $\beta$, such that the phase averaging does not
reduce the visibility. This behaviour is shown in Fig. 2. The
magnitude of the revival is reduced by any decoherence of the
mirror.

The revival demonstrates the coherence of the superposition state
that exists at intermediate times. For $\kappa^2 \gtrsim 1$ the
state of the system is a superposition involving two distinct
mirror positions. More precisely, for a thermal mirror, the state
of the system is a mixture of such superpositions. However, this
does not affect the fundamentally non-classical character of the
state. We now discuss the experimental requirements for achieving
such a superposition and observing its recoherence at $t=2
\pi/\omega_m$.

\begin{figure}
\includegraphics[width= 0.9 \columnwidth]{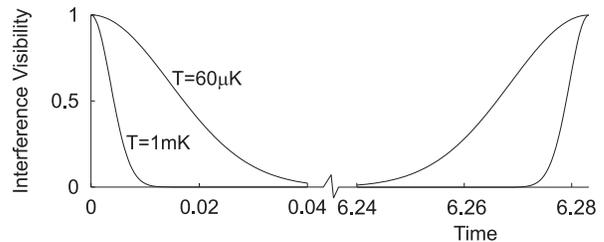}
\caption{Time evolution of the interference visibility of the
photon over one period of the mirror's motion for $\kappa=1$,
$T=1$mK and $T=60 \mu$K. The time is given in units of
$1/\omega_m$. The visibility decays very fast after $t=0$, but in
the absence of decoherence there is a sharp revival of the
visibility after a full period. The width of each peak scales like
$1/\sqrt{T}$. } \label{peak}
\end{figure}

Firstly, we require $\kappa^2 \gtrsim 1$, which physically means
that the momentum kick imparted by the photon has to be larger
than the initial quantum uncertainty of the mirror's momentum. Let
$N$ denote the number of roundtrips of the photon in the cavity
during one period of the mirror's motion, such that
$2NL/c=2\pi/\omega_m$. This allows us to rewrite the condition
$\kappa^2 \gtrsim 1$ as \be \frac{2 \hbar N^3 L}{\pi c M \lambda^2
} \gtrsim 1, \label{required} \ee where $\lambda$ is the
wavelength of the light. The factors entering Eq. (\ref{required})
are not all independent. The achievable $N$, which is determined
by the quality of the mirrors, and the minimum mirror size (and
hence $M$) both depend strongly on $\lambda$. The mirror's lateral
dimension should be an order of magnitude larger than $\lambda$ to
limit diffraction losses. Its thickness in order to achieve
sufficiently high reflectivity depends on $\lambda$ as well.

Eq. (\ref{required}) allows one to compare the viability of
different wavelength ranges. While the highest values for $N$ are
achievable for microwaves (up to $10^{10}$), this is counteracted
by their cm wavelengths. On the other hand there
are no good mirrors for highly energetic photons. The optical
regime seems optimal. Here we propose an experiment with $\lambda$
around 630 nm.

The cavity mode needs to have a very narrow focus on the tiny
mirror, which requires the other cavity end mirror to be large due
to beam divergence. The maximum cavity length is therefore limited
by the difficulty of making large high quality mirrors. We propose
a cavity length of 5 cm, and a small mirror size of $10 \times 10 \times 10$
microns, leading to a mass of order $5 \times 10^{-12}$ kg.

Such a small mirror on a mechanical oscillator can be fabricated
by coating a silicon cantilever with alternating layers of
SiO$_{2}$ and a metal oxide. The best current mirrors are made in
this way. A larger silicon oscillator has been coated with
SiO$_2$/Ta$_2$O$_5$ and used as part of a high-finesse cavity in
Ref. \cite{schiller}.

For the above dimensions the condition (\ref{required}) is
satisfied for $N=5.6 \times 10^6$. Correspondingly, we are aiming
for a photon loss per reflection not larger than $3 \times
10^{-7}$, about a factor of 4 below the best reported values for
such mirrors \cite{rempe}, and for a transmission of $10^{-7}$,
consistent with the $10 \mu$m mirror thickness \cite{hood}. For
these values, about 1\% of the photons are still left in the
cavity after a full period of the mirror. For the above values of
$N$ and $L$ one obtains a frequency $\omega_m=2 \pi \times 500$
Hz. This leads to a quantum uncertainty for the mirror position of
order $10^{-13}$ m, which corresponds to the necessary
displacement in the superposition in order to achieve
$\kappa^2\sim 1$.

The fact that a relatively large $L$ is needed to satisfy Eq.
(\ref{required}) implies that the creation of superpositions
following the related proposal of Ref. \cite{bose}, which relies
on a micro-cavity, imposes requirements that are probably
beyond the reach of current technology. A large $L$ is
helpful because, for a given $N$, it allows one to use a lower
frequency $\omega_m$, and thus a more weakly bound mirror that is
easier to displace by the photon.

Secondly, the requirement of observing the revival puts a bound on
the acceptable environmental decoherence. To estimate the expected
decoherence we model the mirror's environment by an (Ohmic) bath
of harmonic oscillators. The effect of this environment can
approximately be described by a decoherence rate
$\gamma_D=\gamma_m k T M (\Delta x)^2/\hbar^2$ governing the decay
of off-diagonal elements between different mirror positions
\cite{decoherence}. Here $\gamma_m$ is the damping rate for the
mechanical oscillator, $T$ is the temperature of the environment,
which is constituted mainly by the internal degrees of freedom of
the mirror and cantilever, and $\Delta x$ is the separation of two
coherent states that are originally in a superposition. This
approximation is strictly valid only for times much longer than
$2\pi/\omega_m$  and for $\Delta x$ large compared to the width of
the individual wavepackets. A more rigorous description would
follow Ref. \cite{strunz} and take into account the movement of
the mirror. Our analysis indicates that the order of magnitude of
the decoherence is well captured by $\gamma_D$. Assuming that the
experiment achieves $\kappa^2 \gtrsim 1$, i.e. a separation by the
size of a coherent state wavepacket, $\Delta x \sim
\sqrt{\frac{\hbar}{M\omega_m}}$, the condition $\gamma_D \lesssim
\omega_m$  can be cast in the form \be Q \gtrsim \frac{kT}{\hbar
\omega_m}=\bar{n}, \ee where $Q=\omega_m/\gamma_m$ is the quality
factor of the mechanical oscillator. Bearing in mind that
$Q\gtrsim 10^5$ has been achieved \cite{mamin} for silicon
cantilevers of approximately the right dimensions and frequency,
this implies that the temperature has to be 3 mK or less.

Thirdly, the stability requirements for the experiment are very
strict. The phase of the interferometer has to be stable over the
whole measurement time. This means that the distance between the
large cavity end mirror and the equilibrium position
\cite{thermal} of the small mirror has to be stable to of order
$\lambda/20 N=0.6 \times 10^{-14}$m. The required measurement time
can be determined in the following way. A single run of the
experiment starts by sending a weak pulse into the interferometer,
such that on average 0.1 photons go into either cavity. This
probabilistically prepares a single-photon state as required to a
good approximation. The two-photon contribution has to be kept low
because it causes noise in the interferometer. From Eq.
(\ref{thermal}) the width of the revival peak is $2/\kappa
\omega_m \sqrt{\bar{n}}$. This implies that only a fraction $\sim
1/\pi\sqrt{\bar{n}}$ of the remaining photons will leak out in the
time interval of the revival. It is therefore important to work at
the lowest possible temperature. A temperature of 60 $\mu$K has
been achieved with a nuclear demagnetization cryostat \cite{yao}.

Considering the required low value of $\omega_m$ and the fact that
approximately 1\% of the photons remain after a full period for
the assumed loss, this implies a detection rate of approximately
100 photons per hour in the revival interval, given a detection
efficiency of 70 \%. Thus a measurement time of order $10$ minutes
should give statistically significant results. After one such
measurement period the interferometer can be readjusted, and the
experiment can be repeated. Stability of order $10^{-13}$m/min for
an STM inside at 8 K was achieved with a rather simple suspension
\cite{stipe}. Gravitational wave observatories using
interferometers also require very high stability in order to have
an length sensitivity of $10^{-19}$m over timescales of a ms or
greater \cite{ligo}, for arm lengths of order $1$ km.

The experiment requires ultra-high vacuum conditions in order to
ensure that events where an atom hits the cantilever are
sufficiently rare not to cause significant errors, which is at the
level of about 5/s. Background gas particle densities of order
100/cm$^3$ have been achieved \cite{gabrielse} and are sufficient
for our purposes.

After every single run of the experiment the mirror has to be
damped to reset it to its initial (thermal) state. This could be
done by electric or magnetic fields, e.g. following Ref.
\cite{damping}, where a Nickel sphere was attached to the
cantilever, whose $Q$ could then be changed by three orders of
magnitude by applying a magnetic field.

There are several ways in which the total measurement time could
be decreased, thus relaxing the stability requirements. Switchable
cavity mirrors would allow coupling the photon into and out of the
cavity at any desired time, instead of relying on postselection.
This allows the data collection in the revival interval to be
increased by three orders of magnitude. The transmission of a
distributed Bragg reflector (DBR) can be changed by altering the
refractive index of one of the two materials. Highly reflective
mirrors have been fabricated from alternating layers of GaAs and
AlAs \cite{sale}. The refractive index of GaAs can be changed
dramatically by optical pumping \cite{huang}. A change from $3.6$
to $3.15$ on a ps timescale was observed at a probe wavelength of
830nm for pump intensities in the range $1$ kJ/m$^{2}$ (just below
the damage threshold of the material). Such a large change in
refractive index would allow efficient optical switching
\cite{scalora} and is a promising avenue that deserves
experimental investigation. The main open question is whether the
absorption of the materials \cite{sturge} can be made low enough
to match our very strict demands.

Since the width of the revival peak scales like $1/\sqrt{T}$, the
required measurement time can also be decreased by decreasing the
temperature below 60 $\mu$K. Passive cooling techniques may be
improved. In addition, active cooling of mirror oscillators has
been proposed \cite{mancinicooling}, and even implemented
experimentally for a large mirror \cite{cohadon}. The mirror's
movement is sensed via the output light of an optical cavity, and
a feedback loop is used to damp the motion. Applying the theory of
Ref. \cite{mancinicooling} to our mirror, one finds that cooling
even to the ground state of the center of mass motion is
conceivable. This would reduce the required measurement time, and
thus the stability requirements, by a factor of 50.

In principle the proposed setup has the potential to test wave
function reduction models, in particular the one of Ref.
\cite{penrose}. Based on \cite{penrose} and \cite{strunz}, we
estimate that an improvement of the ratio $Q/T$ by about five
orders of magnitude from the values discussed in this paper
($Q=10^5$ and $T=60 \mu$K) would make the predicted wavefunction
decoherence rate comparable to the environmental decoherence rate.
Improvements in $Q$ are certainly conceivable. In particular, $Q$
is known to increase with decreasing temperature \cite{mamin}.
Active cooling methods for the mirror's center of mass motion as
described above could in principle also be used to indirectly cool
the mirror's internal degrees of freedom. Whether such sympathetic
cooling is a realistic avenue is in itself a fascinating subject
for further study.

We have performed a detailed study of the experimental
requirements for the creation and observation of quantum
superposition states of a mirror consisting of $10^{14}$ atoms,
approximately nine orders of magnitude more massive than any
superposition observed to date. Our analysis suggests that, while
very demanding, this goal appears to be in reach of current
technology. It is remarkable that a tabletop experiment has the
potential to test quantum mechanics in an entirely new regime.
Preliminary experiments on components of the proposal are
currently under way.

This work was supported by the E.U. (IST$-1999-10033$). W.M. is
supported by EPSRC (award no. 00309297). C.S. is supported by a
Marie Curie Fellowship of the E.U. (no. HPMF-CT-2001-01205). R.P.
thanks the NSF for support under contract 00-90091 and the
Leverhulme Foundation for an Emeritus Fellowship. We would like to
thank S. Bose, M. Davies, R. Epstein, T. Knuuttila, R. Lalezari,
J. Pethica, and J. Roberts for useful discussions.


\begin{thebibliography}{99}
\bibitem{schroedinger} E. Schr\"{o}dinger, Die Naturwissenschaften {\bf 23}, 807 (1935).
\bibitem{decoherence} D. Giulini {\it et al.}, {\it Decoherence and the Appearance of a Classical World in
Quantum Theory} (Springer, Berlin, 1996); W.H. Zurek, Phys. Today
{\bf 44}, 36 (1991).
\bibitem{bec} J. Ruostekoski, M. J. Collett, R. Graham, and D. F. Walls, Phys. Rev. A {\bf 57},
511 (1998); J. I. Cirac, M. Lewenstein, K. Molmer, and P. Zoller,
Phys. Rev. A {\bf 57}, 1208 (1998).
\bibitem{bose} S. Bose, K. Jacobs, and P.L. Knight, Phys. Rev. A {\bf 59}, 3204 (1999).
\bibitem{armour} A.D. Armour, M.P. Blencowe, and K.C. Schwab, Phys. Rev. Lett. {\bf 88}, 148301 (2002).
\bibitem{squidexp} C.H. van der Wal {\it et al.}, Science {\bf 290}, 773 (2000);
J.R. Friedman, V. Patel, W. Chen, S.K. Tolpygo, and J.E. Lukens,
Nature {\bf 406}, 43 (2000).
\bibitem{c60} M. Arndt {\it et al.}, Nature {\bf 401}, 680 (1999).
\bibitem{collapse} G.C. Ghirardi, A. Rimini and T. Weber, Phys.
Rev. D {\bf 34}, 470 (1986); G.C. Ghirardi, P. Pearle, and A.
Rimini, Phys. Rev. A {\bf 42}, 78 (1990); I.C. Percival, Proc.
Roy. Soc. London, Ser. A {\bf 447}, 189 (1994); D.I. Fivel, Phys.
Rev. A {\bf 56}, 146 (1997); L. Diosi, Phys. Rev. A {\bf 40}, 1165 (1989).

\bibitem{penrose} R. Penrose, in A. Fokas {\it et al.} (Eds.), {\it
Mathematical Physics 2000} (Imperial College, London, 2000).
\bibitem{schmiedmayer} J. Schmiedmayer, R. Penrose, D. Bouwmeester, J. Dapprich, H. Weinfurter, A. Zeilinger, private communication (1997).
See also D. Bouwmeester {\it et al.}, in N. Dadhich and J.
Narlikar (Eds.), {\it Gravitation and Relativity: At the turn of
the Millennium} (IUCAA, Pune, 1998).
\bibitem{law} C.K. Law, Phys. Rev. A {\bf 51}, 2537 (1994); C.K. Law, Phys. Rev. A {\bf 49}, 433
(1993).
\bibitem{mancini} S. Mancini, V.I. Man'ko, and P. Tombesi, Phys. Rev. A {\bf 55}, 3042
(1997); S. Bose, K. Jacobs, and P.L. Knight, Phys. Rev. A {\bf
56}, 4175 (1997).

\bibitem{schiller} I. Tittonen {\it et al.}, Phys. Rev. A {\bf
59}, 1038 (1999).

\bibitem{rempe} G. Rempe, R.J. Thompson, H.J. Kimble, and R.
Lalezari, Opt. Lett. {\bf 17}, 363 (1992).
\bibitem{hood} C.J. Hood, H.J. Kimble, and J. Ye, Phys. Rev. A
{\bf 64}, 033804 (2001).
\bibitem{strunz} W.T. Strunz and F. Haake, quant-ph/0205108 (2002)
\bibitem{mamin} H.J. Mamin and D. Rugar, Appl. Phys. Lett. {\bf
79}, 3358 (2001).

\bibitem{thermal} As we have seen above, the thermal
fluctuations of the small mirror around this equilibrium position
(which are of order $10^{-12}$ m for 60 $\mu$K) lead to a
narrowing of the visibility peaks, but do not destroy the revival.

\bibitem{yao} W. Yao {\it et al.}, J. Low Temp. Phys. {\bf
120}, 121 (2000).

\bibitem{stipe} B.C. Stipe, M.A. Rezaei and W. Ho., Rev. Sci. Inst. {\bf 70}, 137
(1999).

\bibitem{ligo} S. Rowan and J. Hough, Living Review in Relativity, {\bf 3}, 2000-3 (2000),

\bibitem{gabrielse} G. Gabrielse {\it et al.}, Phys. Rev. Lett. {\bf 65}, 1317 (1990).

\bibitem{damping} K. Wago, D. Botkin, C.S. Yannoni, and D. Rugar,
Appl. Phys. Lett. {\bf 72}, 2757 (1998).


\bibitem{sale} T.E. Sale, {\it Vertical Cavity Surface Emitting Lasers} (Research Studies Press, Baldock, 1995).

\bibitem{huang} L. Huang, J.P. Callan, E.N. Glezer, and E. Mazur, Phys. Rev. Lett. {\bf 80}, 185 (1998).

\bibitem{scalora} M. Scalora, J.P. Dowling, C.M. Bowden, and M.J.
Bloemer, Phys. Rev. Lett. {\bf 73}, 1368 (1993).

\bibitem{sturge} M.D. Sturge, Phys. Rev. {\bf 127}, 768 (1962).

\bibitem{mancinicooling} S. Mancini, D. Vitali, and P. Tombesi,
Phys. Rev. Lett. {\bf 80}, 688 (1998).
\bibitem{cohadon} P.F. Cohadon, A. Heidmann, and M. Pinard,
Phys. Rev. Lett. {\bf 83}, 3174 (1999).



\end{thebibliography}
\end{document}